\documentstyle[preprint,aps]{revtex}
\tightenlines
\input  epsf.tex

\begin{document}
\title{M\o ller energy of the nonstatic spherically symmetric metrics.}
\author{S.~S.~Xulu\thanks{%
E-mail: ssxulu@pan.uzulu.ac.za}}
\address{Department of Applied Mathematics, University of Zululand,\\
Private Bag X1001, 3886 Kwa-Dlangezwa, South Africa}
\maketitle
%********************************************************

\begin{abstract}
The energy distribution in  the most general nonstatic spherically symmetric
space-time is obtained using  M\o ller's energy-momentum complex. This result
is compared with the energy expression obtained by using  the energy-momentum 
complex of Einstein. Some examples of energy distributions in different 
prescriptions are  discussed.
\end{abstract}
%********************************************************


\section{Introduction} \label{sec:intro}

The absence of a unique way of defining energy and momentum in general 
relativity has caused much debate. This subject  continues to be one of the 
most active areas of research in general relativity. In spite of  many 
attempts aimed at obtaining an adequate  expression for local or quasi-local
energy and momentum, there is still no generally accepted definition known. 
This has resulted in diverse viewpoints on this subject. In a series of 
papers, Cooperstock \cite{Coop} hypothesized  that in general 
relativity energy and momentum are  located  only  to  the regions of 
nonvanishing energy-momentum tensor and consequently the gravitational waves
are not carriers of energy and momentum. This hypothesis has neither been
proved nor disproved.  Since the advent of Einstein's energy-momentum complex, 
used for calculating energy and momentum  in a general relativistic system, 
many  energy-momentum complexes have been found, for instance,  Landau and
Lifshitz, Papapetrou,  and Weinberg (see in \cite{KSV99} and also
references therein). The 
major difficulty with these definitions  is that they are coordinate-dependent,
with the computations of energy and momentum only giving meaningful results if
calculations are carried out in ``Cartesian coordinates''. This motivated 
M\o ller \cite{Moller58} to construct an expression which enables one to
evaluate energy in any coordinate system. However, M\o ller's energy-momentum 
complex suffered  some criticism (see \cite{CritMol}).

Over the past two decades a large number of definitions of quasi-local mass
(associated with a closed two-surface) have been proposed (see in \cite
{BroYor,Hayw94} and references therein).  Though Penrose\cite{Penrose}
pointed out that a  quasi-local mass is  conceptually important, a serious
problem with the known quasi-local mass definitions is  that these do not
comply even for the Reissner-Nordstr\"{o}m and Kerr space-times\cite
{Bergqv}. Moreover, the seminal quasi-local mass definition of Penrose is not
adequate to handle the Kerr metric\cite{BT}. On the contrary, several 
energy-momentum complexes have been showing a high degree
of consistency in giving the same and acceptable energy and momentum
distribution for a given space-time. This has been found  for many  
asymptotically
flat\cite{Asym,KSV92,V90bCpRi,KSV97,ChaVir96}
as well as asymptotically non-flat 
space-times\cite{NonAsym}.

Recently Virbhadra\cite{KSV99} investigated whether or not the energy-momentum
complexes of Einstein, Landau and Lifshitz, Papapetrou, and Weinberg give the
same energy distribution for the most general nonstatic spherically symmetric
metric, and to a great surprise he found  that  these definitions disagree.
He noted that the energy-momentum complex of Einstein furnished a consistent
result for the Schwarzschild metric whether one calculates in Kerr-Schild Cartesian
coordinates or Schwarzschild Cartesian coordinates.  The definitions
of LL, Papapetrou and Weinberg give the same result as in the Einstein 
prescription if computations are performed in Kerr-Schild Cartesian 
coordinates; however, they disagree with the Einstein definition if computations
are done in Schwarzschild Cartesian coordinates. Thus, the definitions of
LL, Papapetrou and Weinberg do not furnish a consistent result. Based on
this and some other investigations, Virbhadra remarked that the Einstein
method  is the best among all known (including quasi-local mass definitions)
for energy distribution in a space-time. 
Recently  in an important paper Lessner\cite{Lessner} argued that the M\o ller
energy-momentum expression   is a powerful concept of energy and momentum
in general relativity. So in the present paper we wish  to revisit the 
M\o ller energy-momentum expression and use it to compute the energy
distribution in  the most general nonstatic spherically symmetric space-time
and compare this result with  one obtained by Virbhadra in the Einstein
prescription. 
Throughout this paper we use $G=1,c=1$ units and follow the convention that 
Latin indices take values from $0$ to $3$ and Greek indices take values from 
$1$ to $3$.

%********************************************************
\section{Virbhadra's result in the Einstein Prescription}

Virbhadra\cite{KSV99} explored the energy distribution in the most general nonstatic
spherically symmetric space-time. He used the energy-momentum complex of
Einstein. The most general nonstatic spherically symmetric space-time
is described by the line element
\begin{equation}
ds^{2}=\alpha (r,t)\,dt^{2}-\beta (r,t)\,dr^{2}-2\gamma (r,t)\,dt\,dr-\sigma
(r,t)\,r^{2}\,(d\theta ^{2}+\sin ^{2}\theta \,d\phi ^{2}) \text{.}
\label{Genmetric}
\end{equation}
This has, amongst others,  the following well-known space-times
 as special cases: The Schwarzschild metric, Reissner-Nordstr\"{o}m
metric, Vaidya metric,  Janis-Newman-Winicour metric,
 Garfinkle-Horowitz-Strominger metric, a general non-static spherically 
symmetric metric of the Kerr-Schild class (discussed  in  Virbhadra's
 paper\cite{KSV99}).

The  Einstein energy-momentum complex is given as
\begin{equation}
\Theta_i{}^{k} = \frac{1}{16 \pi} H^{\ kl}_{I \ \ ,l} ,
\end{equation}
%%%%%%%
where
%%%%%%
\begin{equation}
H_i^{\ kl}  =  - H_i^{\ lk}\ =\  \frac{g_{in}}{\sqrt{-g}}
         \left[-g \left( g^{kn} g^{lm} - g^{ln} g^{km}\right)\right]_{,m} \ .
\end{equation}
%%%%%%%%%%%
$\Theta_0^{\ 0}$ and $\Theta_{\alpha}^{\ 0}$ denote for the energy and momentum
density components, respectively. (Virbhadra\cite{KSV99} mentioned that 
though the  energy-momentum complex found by Tolman differs in form from 
the Einstein energy-momentum complex, both are equivalent in import.) 
 The energy-momentum components 
are expressed by
\begin{equation}
P_i  =  \int \int \int \Theta_i^{\ 0} dx^1  dx^2 dx^3 .
\end{equation}
Further Gauss's theorem  furnishes
\begin{equation}
P_i  = \frac{1}{16 \pi} \ \int\int\ H_i^{\ 0 \alpha} \ \mu_{\beta}\ dS \text{,}
\end{equation}
where $\mu_{\beta}$ is the outward unit normal vector over the infinitesimal 
surface element $dS$.  $P_\alpha$ give momentum  components
$P_1, P_2, P_3$  and $P_{0}$ gives the energy.

Virbhadra transformed the line element $(\ref{Genmetric})$ to ``Cartesian
coordinates'' $(t, x, y, z)$  using $x =  r \sin\theta \cos\phi, 
y = r \sin\theta  \sin\phi,  z =  r \cos\theta$ and $t$ remaining the same
and then used the Einstein energy-momentum complex to obtain the energy 
distribution, which is given below.
\begin{equation}
E_{\rm Einst} =  \frac{r\left[\alpha\left(\beta-\sigma-r \sigma_{,r}\right)
            - \gamma \left(r\sigma_{,t}-\gamma\right)\right]}
           {2 \sqrt{\alpha \beta+\gamma^2}}.
\label{EEinst} 
\end{equation}
In the next Section we obtain the energy distribution for the same metric
in M\o ller's formulation.
%********************************************************

\section{ENERGY DISTRIBUTION IN M\O LLER's FORMULATION}

In this Section we first write the energy-momentum complex of M\o ller
and then use this for the most general nonstatic spherically symmetric
metric given by the equation $(\ref{Genmetric})$. Note that as the M\o ller
complex is not restricted to the use of ``Cartesian coordinates'' we perform
the computations in $t,r,\theta,\phi$ coordinates, because computations
in these coordinates are  easier compared to those in $t,x,y,z$  coordinates.

The following is the M\o ller  energy-momentum complex ${\Im }_{I}^{\ k}$ 
 \cite{Moller58} :
\begin{equation}
{\Im }_{I}^{\ k}=\frac{1}{8\pi }{\chi}_{I\ \ ,l}^{\ kl} \quad \text{,} 
 \label{MollerEMC}
\end{equation}
which satisfy the local conservation laws: 
\begin{equation}
\frac{\partial {\Im }_{I}^{\ k}}{\partial x^{k}}= 0 \text{ .}
\label{ConsvLaws}
\end{equation}
The antisymmetric superpotential ${\chi}_{I}^{\ kl}$ is  
\begin{equation}
{\chi}_{I}^{\ kl} = - {\chi}_{I}^{\ lk} = \sqrt{-g} \left[
g_{in,m}-g_{im,n}\right]  g^{km} g^{nl}\text{ \ .}  
\label{Chi}
\end{equation}
${\Im }_{0}^{\ 0}$ is the energy density and ${\Im }_{\alpha}^{\ 0}$ are
the momentum density components. Obviously, the energy and momentum components
are given by
\begin{equation}
P_{I}=\int \int \int {\Im }_{I}^{\ 0} dx^{1} dx^{2} dx^{3} \text{\ ,}
\end{equation}
where $P_{0}$ is the energy  while $P_{\alpha }$ denote for
the momentum  components. Further  Gauss's theorem  furnishes the energy 
$E$ given by  
\begin{equation}
E = \frac{1}{8\pi } \int \int {\chi}_{0}^{\ 0\beta }\ \mu _{\beta }\ dS
\label{energy}
\end{equation}
where $\mu _{\beta }$ is the outward unit normal vector over an
infinitesimal surface element $dS$.

For the line element $(\ref{Genmetric})$  under consideration we calculate 
\begin{equation}
{\chi}_{0}^{\ 01}=\frac{\left( \alpha _{,r}+\gamma _{,t}\right)
    \sigma r^{2}\sin\theta}{
\left( \alpha \beta +\gamma^{2}\right)^{1/2}} 
 \text{\  ,}
 \label{chi001}
\end{equation}
which is the only required component of ${\chi}_{I}^{kl}$ for our purpose.

Using the above expression in equation  $(\ref{energy})$ we obtain the energy 
distribution 
 \begin{equation}
E_{\rm M\o l} 
= \frac{\left( \alpha _{,r}+\gamma _{,t}\right) \sigma r^2 }{2\left( \alpha
\beta +\gamma ^{2}\right)^{1/2}}\text{ \  .}  
\label{EMol}
\end{equation}
It is evident that the energy distribution for the most general nonstatic
spherically symmetric metric  the definitions of Einstein and M\o ller
disagree in general (compare $(\ref{EEinst})$ with $(\ref{EMol})$).
However, these furnish the same results for some space-times, for instance,
the Schwarzschild and Vaidya space-times\cite{KSV92}.  In the next Section
we will compute energy distribution in a few  space-times using 
$(\ref{EEinst})$ and  $(\ref{EMol})$.
%********************************************************

\section{\protect\bigskip Examples}
In this Section we discuss  a  few examples of space-times in the Einstein 
as well  as the M\o ller prescriptions.

\begin{enumerate}
\item {\bf The Schwarzschild solution}\\
This solution is  expressed by the line element
\begin{equation}
ds^2 = \left(1-\frac{2M}{r}\right) dt^2 
    - \left(1-\frac{2M}{r}\right)^{-1} dr^2
    -r^2 \left(d\theta +\sin^2\theta^2 d\phi^2\right) \text{\ .}
\end{equation}
 Equations $(\ref{EEinst})$ and 
$(\ref{EMol})$ furnish (see also in \cite{Moller58,KSV97})
\begin{equation}
 E_{\rm Einst} = E_{\rm M\o l} = M
\end{equation}
showing that these two definitions of energy distribution agree for the
Schwarzschild space-time.

\item {\bf The Reissner-Nordstr\"{o}m  solution}\\
  The Reissner-Nordstr\"{o}m  solution is given by 
\begin{equation}
ds^2 = \left(1-\frac{2M}{r}+\frac{e^2}{r^2}\right) dt^2 
    - \left(1-\frac{2M}{r}+\frac{e^2}{r^2}\right)^{-1} dr^2
    -r^2 \left(d\theta^2 +\sin^2\theta d\phi^2\right) \text{\ ,}
\end{equation}
and the antisymmetric electromagnetic field tensor
\begin{equation}
F_{tr} = \frac{e}{r^2} \text{\ ,}
\end{equation}
where $M$  and $e$ are respectively the mass and electric charge 
parameters.

For this space-time  equations $(\ref{EEinst})$ and  $(\ref{EMol})$ furnish 
(see also in \cite{V90bCpRi})
\begin{equation}
 E_{\rm Einst}  = M - \frac{e^2}{2r}
\end{equation}
and 
\begin{equation}
 E_{\rm M\o l} = M - \frac{e^2}{r} \text{\ .}
\end{equation}
%Note that the definition of Einstein yield the flat space limit, whereas that
%of M\o ller does not do that (for more discussions see \cite{CoopRi}

\item {\bf The Janis-Newman-Winicour  solution\footnote{
This solution has been usually incorrectly referred to in the literature
 as the Wyman solution.
Virbhadra\cite{KSV97} proved that the Wyman solution is the same as the
Janis-Newman-Winicour solution. As Janis, Newman and Winicour obtained this 
solution much before Wyman, Virbhadra\cite{KSV99} rightly referred to this 
as the  Janis-Newman-Winicour solution.}} \\

This solution is given by
\begin{equation}
ds^2 = \left(1-\frac{B}{r}\right)^{\mu} dt^2 
      - \left(1-\frac{B}{r}\right)^{-\mu} dr^2 
      - \left(1-\frac{B}{r}\right)^{1-\mu}  
           r^2 \left(d\theta^2  +\sin^2\theta \  d\phi^2\right)
\label{JNWLE}
\end{equation}
and the scalar field
\begin{equation}
\Phi = \frac{q}{B\sqrt{4\pi}} \ln\left(1-\frac{B}{r}\right),
\label{JNWPHI}
\end{equation}
where
\begin{eqnarray}
\mu &=& \frac{2M}{B}, \nonumber\\
B &=& 2 \sqrt{M^2+q^2}.
\label{MUB}
\end{eqnarray}
$M$ and $q$ are the mass and scalar charge parameters respectively.
For $q=0$ this solution  furnishes the Schwarzschild solution.

Virbhadra\cite{KSV99} computed the energy expression for this metric using 
the Eq. $(\ref{EEinst})$. We do the same here using equation  $(\ref{EMol})$.
Thus we  find that
\begin{equation}
 E_{\rm Einst} = E_{\rm M\o l} = M \text{,}
\end{equation}
which exhibits  that these two definitions of energy distribution 
agree for the Janis-Newman-Winicour space-time.

\item{\bf Garfinkle-Horowitz-Strominger  solution}

The Garfinkle-Horowitz-Strominger  static spherically symmetric asymptotically
flat solution  is described by  (see in \cite{GHS})
\begin{equation}
ds^{2} = (1 - \frac{r_+}{r })(1 - \frac{r_-}{r })^\nu dt^2 
        - (1 - \frac{r_+}{r })^{-1}(1 - \frac{r_-}{r })^{-\nu} dr^2 
        -  (1 - \frac{r_-}{r })^{1-\nu} r^2 (d \theta^2 + sin^2 \theta 
      d \phi^2) \text{\ ,}
 \end{equation}
 the dilaton field (denoted by $\Phi$) by
\begin{equation}
exp(2\Phi) = \left(1-\frac{r_-}{r}\right)^{\frac{1-\nu}{\lambda}}
\end{equation}
 and the antisymmetric electromagnetic field tensor is
\begin{equation}
F_{ik} = \frac{e}{r^2} \text{\ ,}
\end{equation}
where
\begin{equation}
~~\nu = \frac{1-{\lambda}^2}{1+{\lambda}^2}.
\end{equation}
$r_-$ and $r_+$ are related to mass M and charge e parameters as follows:
\begin{eqnarray}
M &=& \frac{r_+ + \nu r_-}{2},\nonumber\\
e^2  &=& \frac{r_+ r_-}{(1 + \lambda^2)}.
\end{eqnarray}
$\lambda = 0$  in this solution gives the Reissner-Nordstr\"{o}m  solution.
Virbhadra\cite{KSV97} proved that for $e=0$ the Garfinkle-Horowitz-Strominger
solution reduces to the Janis-Newman-Winicour solution; this fact was not 
noticed by Garfinkle, Horowitz and Strominger in their paper.

  Chamorro and Virbhadra\cite{ChaVir96} computed the energy distribution in
Garfinkle-Horowitz-Strominger space-time using the energy-momentum complex of
Einstein. They found that
\begin{equation}
 E_{\rm Einst}  = M - \frac{e^2}{2r}\left(1-\lambda^2\right) \text{\ .}
\label{GHSEE}
\end{equation}
We compute the energy distribution for the Garfinkle-Horowitz-Strominger 
space-time using equation $(\ref{EMol})$ and obtain
\begin{equation}
 E_{\rm M\o l}  = M - \frac{e^2}{r}\left(1-\lambda^2\right) \text{\ .}
\label{GHSME}
\end{equation}
Thus, these two definitions give different results (there is a  difference
of a factor $2$ in the second term).

Now defining 
\begin{equation}
{\cal E}_{\rm Einst} :=  \frac{E_{\rm Einst}}{M},  \quad
{\cal E}_{\rm  M\o l} :=  \frac{E_{\rm M\o l}}{M},  \quad
{\cal Q}  :=    \frac{e}{M},  \quad
{\cal R}  :=   \frac{r}{M}  
\end{equation}
%********************************************************
the equations  $(\ref{GHSEE})$ and  $(\ref{GHSME})$ may be expressed as  
\begin{equation}
{\cal E}_{\rm Einst} =
    1-\frac{{\cal Q}^{2}}{2{\cal R}} \left(1-\lambda^2\right)
\label{CalGHSEE}
\end{equation}
and 
\begin{equation}
{\cal E}_{\rm M\o l} =
         1-\frac{{\cal Q}^{2}}{{\cal R}} \left(1-\lambda^2\right)
\label{CalGHSME}
\end{equation}
For $\lambda^2 = 1, {\cal E}_{\rm Einst} = {\cal E}_{\rm M\o l} = M$;
however, they differ for any other values of $\lambda^2$. For any values of
$\lambda^2 < 1$,  ${\cal E}_{\rm Einst}$  as well as ${\cal E}_{\rm M\o l}$
decrease with an increase in ${\cal Q}^2$ and increase with increase in 
${\cal R}$.
${\cal E}_{\rm Einst} > {\cal E}_{\rm M\o l}$ and they asymptotically
($\cal R \rightarrow \infty$) reach the value $1$.
The situation is just opposite for any values of $\lambda^2 > 1$ :  
${\cal E}_{\rm Einst}$  as well as ${\cal E}_{\rm M\o l}$
increase with an increase in ${\cal Q}^2$ and decrease with increase in $R$.
${\cal E}_{\rm Einst} < {\cal E}_{\rm M\o l}$ and they asymptotically
($\cal R \rightarrow \infty$) reach the value $1$.

We plot the energy distributions ${\cal E}_{\rm Einst}$ and  
${\cal E}_{\rm M\o l}$ for  $\lambda = 0$ 
(Reissner-Nordstr\"{o}m space-time) in the figure 1 and for $\lambda^2 =1.2$ 
in  figure  2.
\end{enumerate}
%********************************************************
\begin{figure*}
\epsfxsize 12cm
\epsfbox{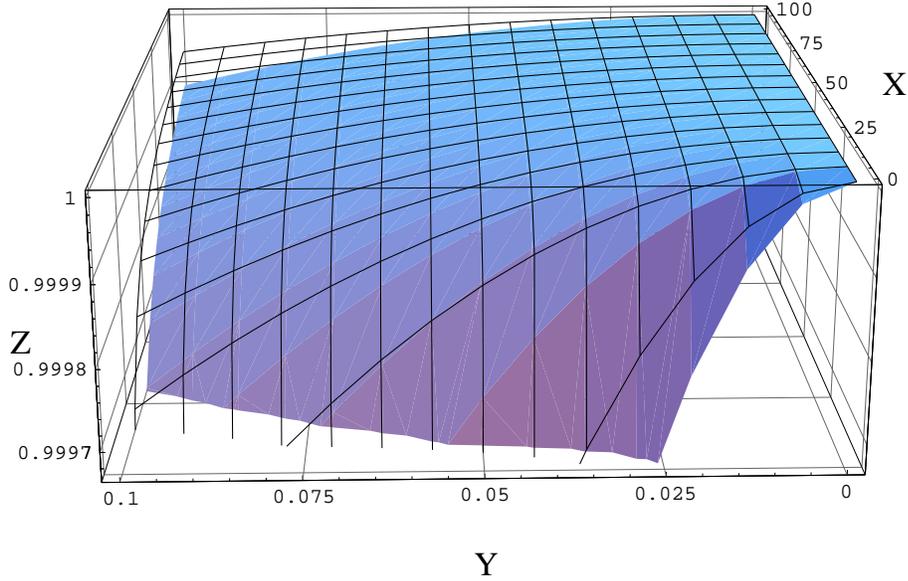}
\caption{
${\cal E}_{\rm Einst}$ and  ${\cal E}_{\rm M\o l}$ on Z-axis are 
    plotted against  ${\cal R}$ on X-axis and ${\cal Q}$ on Y-axis for 
   ${\lambda} = 0$ (Reissner-Nordstr\"{o}m metric). 
  The upper (grid-like) and lower surfaces 
 are for ${\cal E}_{\rm Einst}$ and  ${\cal E}_{\rm M\o l}$ respectively.
}
\label{fig1}
\end{figure*}
%********************************************************
\begin{figure*}
\epsfxsize 12cm
\epsfbox{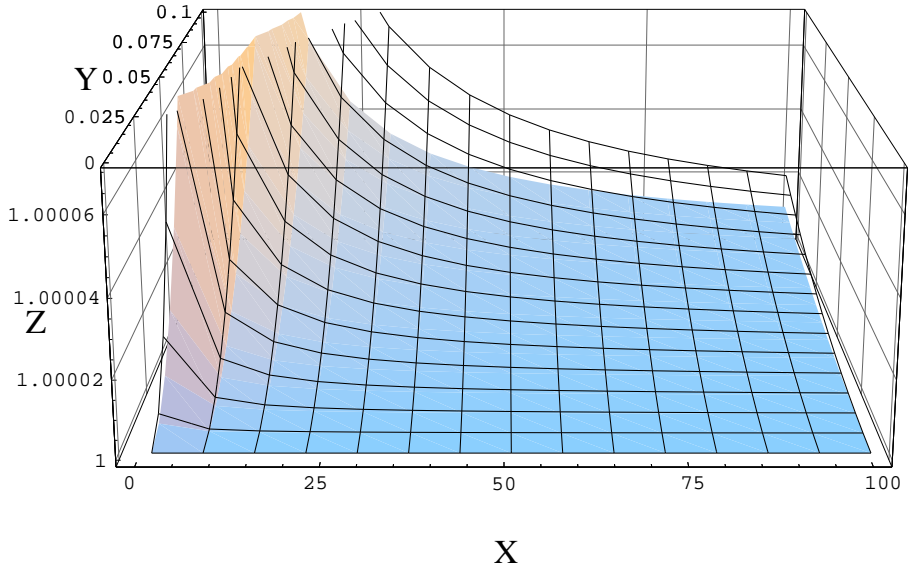}
\caption{
${\cal E}_{\rm Einst}$ and  ${\cal E}_{\rm M\o l}$ on Z-axis are 
    plotted against  ${\cal R}$ on X-axis and ${\cal Q}$ on Y-axis for 
   ${\lambda^2} = 1.2$. 
  The upper (grid-like) and lower surfaces 
 are for ${\cal E}_{\rm M\o l}$  and  ${\cal E}_{\rm Einst}$ respectively.
}
\label{fig2}
\end{figure*}
\newpage
%********************************************************
\section{\protect\bigskip Conclusion}

Based on some  analysis of the results known with many prescriptions
for energy distribution (including some  well-known quasi-local mass 
definitions) in a given space-time Virbhadra\cite{KSV99}
remarked that the  formulation by Einstein is still the best one.
In a recent  paper Lessner\cite{Lessner} argued that the M\o ller
energy-momentum expression is a powerful concept of energy and momentum
in general relativity, which motivated us to study this further.
We obtained the energy distribution for the most general nonstatic
spherically symmetric metric using  M\o ller's   definition. The
result we found differs in general from that  obtained using the
Einstein energy-momentum complex;  these agree for the
Schwarzschild, Vaidya and Janis-Newman-Winicour space-times, but disagree
for the Reissner-Nordstr\"{o}m space-time.
For the Reissner-Nordstr\"{o}m  space-time $E_{\rm Einst} = M - e^2/(2 r)$
(the seminal Penrose quasi-local mass definition also yields  the same result 
agreeing  with linear theory\cite{Tod}) whereas  $E_{\rm M\o l} = 
M - e^2/r$. This question must be considered important. M\o ller's energy-
momentum complex is not constrained to the use of any particular coordinates 
(unlike  the case of the Einstein complex); however it does not furnish expected
result for the Reissner-Nordstr\"{o}m space-time.  We agree with Virbhadra's
remark  that the Einstein energy-momentum complex is still the best tool
for obtaining energy distribution in a given space-time.
%************************************************************
\acknowledgments
I am grateful  to  NRF  for financial support.
%************************************************************

\end{document}